\documentclass[fleqn,usenatbib]{mnras}
\usepackage{newtxtext,newtxmath}
\usepackage{booktabs}
\usepackage{footmisc}
\usepackage[T1]{fontenc}
\DeclareRobustCommand{\VAN}[3]{#2}
\let\VANthebibliography\thebibliography
\def\thebibliography{\DeclareRobustCommand{\VAN}[3]{##3}\VANthebibliography}

\usepackage{graphicx}	
\usepackage{amsmath}	

\defcitealias{canameras21}{C21}
\defcitealias{shu22}{S22}

\usepackage{xcolor}

\title[Comparison of neural network algorithms]{ Systematic comparison of neural networks used in discovering strong gravitational lenses}

\author[More et al.]{Anupreeta More,$^{1,2}$\thanks{E-mail: anupreeta@iucaa.in}
Raoul Ca\~nameras,$^{3,4,5}$
Anton T. Jaelani,$^{6,7}$
Yiping Shu,$^{3}$ 
Yuichiro Ishida,$^{8,9}$\and
Kenneth C. Wong,$^{10,11}$
Kaiki Taro Inoue,$^{12}$
Stefan Schuldt,$^{13,14}$
and
Alessandro Sonnenfeld$^{15,16}$ 
\\
$^{1}$Inter-University Centre for Astronomy and Astrophysics, Ganeshkhind, Pune 411007, India.\\
$^{2}$Kavli Institute for the Physics and Mathematics of the Universe (WPI), University of Tokyo, 5-1-5,Kashiwa, Chiba 277-8583, Japan.\\
$^{3}$Max-Planck-Institut für Astrophysik, Karl-Schwarzschild-Str. 1, D-85748 Garching, Germany \\
$^{4}$Technical University of Munich, TUM School of Natural Sciences, Department of Physics, James-Franck-Str. 1, D-85748 Garching, Germany \\
$^{5}$Aix Marseille Univ, CNRS, CNES, LAM, Marseille, France \\
$^{6}$Astronomy Research Group and Bosscha Observatory, FMIPA, Institut Teknologi Bandung, Jl. Ganesha 10, Bandung 40132, Indonesia\\
$^{7}$U-CoE AI-VLB, Institut Teknologi Bandung, Jl. Ganesha 10, Bandung 40132, Indonesia\\
$^{8}$Department of Astronomy, The University of Tokyo, 7-3-1 Hongo, Bunkyo-ku, Tokyo 113-0033, Japan \\
$^{9}$Department of Earth and Planetary Sciences, Kyushu University, 744 Motooka, Nishi-ku, Fukuoka 819-0395, Japan \\
$^{10}$Research Center for the Early Universe, Graduate School of Science, The University of Tokyo, 7-3-1 Hongo, Bunkyo-ku, Tokyo 113-0033, Japan \\
$^{11}$National Astronomical Observatory of Japan, 2-21-1 Osawa, Mitaka, Tokyo 181-8588, Japan \\
$^{12}$Kindai University 3-4-1 Kowakae Higashi-Osaka, Osaka Japan \\
$^{13}$Dipartimento di Fisica, Universit\`a  degli Studi di Milano, via Celoria 16, I-20133 Milano, Italy \\
$^{14}$INAF - IASF Milano, via A. Corti 12, I-20133 Milano, Italy\\
$^{15}$Department of Astronomy, School of Physics and Astronomy, Shanghai Jiao Tong University, Shanghai 200240, PR China \\
$^{16}$Leiden Observatory, Leiden University, PO Box 9513, 2300 RA Leiden, The Netherlands \\
}

\date{Accepted XXX. Received YYY; in original form ZZZ}

\pubyear{2024}

\begin{document}
\label{firstpage}
\pagerange{\pageref{firstpage}--\pageref{lastpage}}
\maketitle

\begin{abstract}
Efficient algorithms are being developed to search for strong gravitational lens systems owing to increasing large imaging surveys. Neural networks have been successfully used to discover galaxy-scale lens systems in imaging surveys such as the Kilo Degree Survey, Hyper-Suprime Cam (HSC) Survey and Dark Energy Survey over the last few years. Thus, it has become imperative to understand how some of these networks compare, their strengths and the role of the training datasets as most of the networks make use of supervised learning algorithms. In this work, we present the first-of-its-kind systematic comparison and benchmarking of networks from four teams that have analysed the HSC Survey data. Each team has designed their training samples and developed neural networks independently but coordinated apriori in reserving specific datasets strictly for test purposes.  The test sample consists of mock lenses, real (candidate) lenses and real non-lenses gathered from various sources to benchmark and characterise the performance of each of the network. While each team's network performed much better on their own constructed test samples compared to those from others, all networks performed comparable on the test sample with real (candidate) lenses and non-lenses. We also investigate the impact of swapping the training samples amongst the teams while retaining the same network architecture. We find that this resulted in improved performance for some networks. These results have direct implications on measures to be taken for lens searches with upcoming imaging surveys such as the Rubin-Legacy Survey of Space and Time, Roman and Euclid. 
\end{abstract}


\section{Introduction}
Machine learning applications in astronomy have been growing within the last
decade including the field of gravitational lensing.  In strong gravitational
lensing, multiple lensed images of the same distant galaxy or a quasar are
observed owing to the gravitational deflection by a massive galaxy or a cluster
in the foreground. Since this requires sufficient line-of-sight alignment
between the distant source and the foreground lens with the observer, such lens
systems are rare occurrence. However, with increasing number of large imaging
surveys with sufficiently deep observations, discovery of large lens samples
has become feasible, for instance, from the Dark Energy Survey \citep{Diehl2017,odonnell22}, Survey of Gravitationally-lensed Objects in HSC Imaging \citep[SuGOHI, e.g.,][]{Son18,sonnenfeld20,Jae21,Wong22,Chan23}, Kilo Degree Survey \citep[KiDS, e.g.,][]{petrillo17,Khramtsov2019,Li2020} and DECam Legacy Survey \citep[DECaLS, e.g.,][]{Huang2020,Storfer2022}. 
Searching for lens systems is a classical
pattern-recognition problem as it involves identifying specific configurations,
morphologies and colors that are expected as a result of lensing. Additionally,
the rarity of the lens systems requires sifting through hundreds of images
before a promising candidate lens system is discovered. Thus, this is an apt
challenge that can be addressed with machine learning algorithms.

Supervised, deep learning algorithms based on Convolutional Neural Networks
(CNNs) are favorable as the majority of astronomy data include analysis of multi-wavelength imaging.  In the last few years, the CNNs have been successfully implemented for
searching, primarily, galaxy-scale lenses \citep[e.g.][]{Jacobs17,petrillo17,petrillo19,canameras20,He2020,Rojas2023}.  A few studies have attempted to make a comparison between different neural network algorithms with other lens search methods with real
survey data. For instance, \cite{Jacobs17} compared the results of a CNN search on
Canada-France-Hawaii Telescope Legacy Survey (CFHTLS) data to the results from
a purely visual-inspection based search conducted via Space Warps \citep{Marshall2016,More16}, a citizen science program. It is worth noting that the
Space Warps results from CFHTLS data are also produced using a
supervised-learning approach. Similarly, \cite{More16}, citizen-science-based results are also compared
with non-machine-learning algorithms \citep{Gavazzi2014,More2012}.  Such
comparison studies have suggested that each of these approaches and algorithms
tend to find a subset of lens systems with some overlap with each other.

Others have compared diverse lens search methods which include pure visual inspection and algorithms with/without machine learning on simulated space-based and ground-based datasets \citep{metcalf19}. They highlighted that  multi-band imaging plays an important role in increasing the efficiency of lens identification. Further study by \cite{magro2021} on the same datasets but after applying modified data pre-processing and augmentation showed an improved performance of the various neural networks and emphasised the adaptability of CNNs. In \cite{Knabel2020}, lens search methods such as machine learning, visual inspection and spectroscopy are compared by analysing  the data from the KiDS - Galaxy Mass Assembly (GAMA). They find that each of the methods had distinct selection functions resulting into hardly any overlapping candidates in spite of analysing the same footprints across three different fields. Surveys from upcoming telescopes such as Vera Rubin Observatory\footnote{\url{https://www.lsst.org}}, 
Euclid\footnote{\url{https://sci.esa.int/web/euclid}} and Nancy Grace Roman\footnote{\url{https://roman.gsfc.nasa.gov}} will increase
the rate of detection of lenses by an order of magnitude. The need for efficient
and robust machine learning algorithms is stronger ever than before given the challenge of big data.

In this work, we attempt to do the first systematic comparison of multiple
networks and training sets which are tested on a common and diverse test
dataset. Such a study is crucial in identifying the strengths and weaknesses of
the network architectures along with construction strategies of different
training-validation datasets and thus enabling the development of a superior
and robust approach that will produce lens searches with high efficiency. In \cite{Holloway2024}, a companion study, we combine different machine learning networks and Space Warps with the goal of constructing a unified, superior ensemble classifier that will be much more efficient than any of the individual methods.

The paper is structured as follows. In Sec~\ref{sec:net}, we briefly introduce the
various networks and methodologies used in generating the training-validation
datasets. In Sec~\ref{sec:comm}, we describe the construction of the various common test
datasets. In Sec~\ref{sec:metrics}, we list the metrics used in our comparison study. In Sec~\ref{sec:results}, we present the results and give conclusions in Sec~\ref{sec:conc}.

\section{Overview of neural networks}
\label{sec:net}
Below we give a brief overview of the different neural networks that are used for comparison in this work. The participating teams have used the data from the HSC SSP Public Data Release 2 (PDR2) \citep{Aihara+19} for this study.

\subsection{Canameras et al.}
The classification in \citet[][hereafter C21]{canameras21} uses a residual neural network (ResNet) inspired from the ResNet-18 
architecture \citep{he16}. After the 64\,$\times$\,64 $\times$\,3 input layer, it comprises a total of 18 layers, starting 
with a convolutional layer with 3\,$\times$\,3 convolutional kernels and 64 feature maps, followed by eight residual blocks, 
an average pooling layer, a flattening layer, and closed by a fully connected layer with 16 neurons, and the last 
single-neuron output with sigmoid activation. Each residual block comprises two convolutional layers with 3\,$\times$\,3
kernel sizes and ${\rm stride = 1}$ or 2, batch normalization and nonlinear ReLU activations. Convolutional layers within 
these blocks have 64, 128, 256, and 512 feature maps, respectively.

The network was trained and validated on $gri$ images of the HSC Survey, augmented with small random shifts ranging between -5 
and +5 pixels, and square root stretch (after clipping negative pixels to zero), resulting in a balanced dataset of 40,000 mock
lenses and 40,000 non-lens galaxies. The optimization was performed with mini-batch gradient descent and we used a batch size of 
128 images, a learning  rate of 0.0006, a weight decay of 0.001, and a momentum fixed to 0.9. The binary cross-entropy loss was 
computed over the training and validation sets at each epoch, and we used early stopping to save the best model at minimal 
validation loss.

In C21, this ResNet was chosen among a range of networks to optimize lens identification over all extended galaxies in DR2 with
$i$-band Kron radius $\geq$0.8\arcsec, and without photometric pre-selection. It was tested on sets of 202 grade-A or B 
galaxy-scale lens candidates from SuGOHI, and 91,000 nonlens galaxies in the COSMOS field, with both sets restricted to Kron 
radii $\geq$0.8\arcsec. This specific network, and the score threshold of 0.1 were chosen to reach contamination rates as low 
as 0.01\% while ensuring a recall $>$50\% over the SuGOHI test sample. The results from C21 illustrate the ability of this 
network to efficiently select new strong lens candidates from an extended input sample of 62.5 million galaxies, with moderate 
visual inspection. Output scores tend to shift to higher values in regions with seeing FWHMs simultaneously higher in $r$ band 
and lower in $i$ band, as found over the GAMA09H field. This seeing dependence is discussed in more details in \citet{canameras23}.

\subsection{Shu et al.}
Two lens classifiers were presented in \citet[][hereafter S22]{shu22}, both of which were constructed based on the deep residual network, \textsc{deeplens\_classifier}, pre-built in the \textsc{CMU DeepLens} package \citep{Lanusse18}. The main difference between those two lens classifiers was the mock lens population in the training set. For Classifier-1, the mock lenses in the training set covered a lens redshift range of 0--1.0 with a peak at 
$\approx 0.55$. For Classifier-2, the lens redshift distribution was relatively uniform from 0.4 to 1.0. 
It was shown by S22 that, as a result of the different choices of the training set, Classifier-1 delivered an overall high recall for strong-lens systems up to lens redshift of $\sim 0.8$, while Classifier-2 was more optimised for discovering strong-lens systems with high-redshift ($z \gtrsim 0.7$) lens galaxies. As the strong-lens systems used in this work span a wide lens redshift range, we only consider Classifier-1 from S22 in the following analyses. 

A full description of how Classifier-1 was built, trained, and tested can be found in S22. Here we only summarise a few aspects that are relevant for comparing with the other networks. Classifier-1 was trained and validated on HSC $gri$ images of 43,500 mock lenses and 43,500 non-lens objects. The mock lenses were created in the same way as in C21 and were therefore qualitatively similar to the mock lenses used for training the network in C21. The non-lens objects were from a random subset of the parent sample that Classifier-1 was eventually applied to. Since the key motivation of S22 was to search for strong-lens systems with high-redshift lens galaxies, the parent sample was selected to contain relatively red galaxies using the $g-r$ and $g-i$ colours. There was no cut on the Kron radius, and in fact about two thirds of the parent sample had $i$-band Kron radius smaller than 0.8\arcsec. 

Classifier-1 was optimised based on a test set consisting of 92 grade-A or B strong-lens candidates from the SuGOHI project that were also in the parent sample and 50,000 non-lens objects randomly selected from the parent sample. In S22, the probability threshold was chosen to be $p_{\rm thresh} = 0.9731$, which corresponded to a TPR of 0.85 and an FPR of 0.001 on the test set.  

\subsection{Jaelani et al.} 
The lens classification in J23 uses a classical convolutional neural network (CNN) inspired from the CNN architecture used in \citet{Jacobs17}. The network comprises five convolutional layers with 11 $\times$ 11, 7 $\times$ 7, 5 $\times$ 5, 5 $\times$ 5, and 3 $\times$ 3 kernel sizes; and 64, 128, 128, 256, and 256 filters, respectively. It is followed by four fully connected hidden layers with 1024, 1024, 512, and 512 neurons, and a single-neuron output layer with sigmoid activation. Three Max-pooling layers with 2 $\times$ 2 kernel sizes and stride = 2 are inserted in between the convolutional layers and are essential to make the CNN invariant to local translations of the relevant features in $gri$ image cutouts, while reducing the network parameters. Five dropout regularizations are performed in between convolutional and fully-connected layers to reduce the chance of overfitting by randomly dropping a 0.2 of the output neurons during training with ReLU nonlinear activations. 

The CNN was trained and validated on HSC $gri$ images of 18,660 mock lenses and 18,660 non-lens objects. The augmentations have been applied to the dataset by following: (i) a random rotation in the range [-30 deg, 30 deg]; (ii) a random resizing zoom\_range in the range [0.8, 1.2]; (iii) a random horizontal flipping; (iv) and a random channel\_shift\_range $= 0.9$. The Adam optimization algorithm was chosen to minimize the cross-entropy error function over training data with a learning rate of 0.00005. The CNN was trained for 52 epochs (with 100 epochs are the maximum allowed) using mini-batch stochastic gradient descent with 128 images per batch. We used early stopping after patience 5 epochs if the network did not give better accuracy or loss.

The parent sample of 2.3 million galaxies that we used in J23 was selected based on criteria on, e.g., multiband magnitudes, stellar mass, star formation rate, extendedness limit, and photometric redshift range.

\subsection{Ishida et al.} 
This strong lens classifier (Ishida et al. 2024 in prep.; hereafter I24), uses a classical CNN architecture.  The CNN is composed of six blocks. Each block consists of two convolutional layers with an equal number of filters and a batch normalization layer. Convolutional layers within these blocks have 32, 64, 64, 64, 128, and 128 filters, respectively, with ReLU activation. The first layer uses a 7x7 kernel for convolution, and subsequent layers use a 3$\times$3 kernel. Three max-pooling layers with a kernel size of 2$\times$2 are inserted in between blocks with different numbers of filters, as well as after the last block.  These are followed by two fully-connected layers with 128 and 64 neurons with ReLU activation,  and a single-neuron output layer with sigmoid activation.  Dropout layers with a dropout rate of 0.4 are inserted between the two fully-connected layers, as well as between the fully-connected and output layer.

The training and validation data are the same as for the J23 network, comprising 18,660 mock lenses and 18,660 non-lenses.  We scale the fits image data using an algorithm (hereafter ``SDSS normalization'') based on \citet{lupton04}.  We first scale the $g$, $r$, and $i$ band images by multiplicative factors of 2.0, 1.2, and 1.0, respectively.  These values were determined through testing of various scaling factors and were found to give the best results.  We then apply the normalization described by the equations

\begin{eqnarray} 
I = \frac{g + r + i}{3} \, \,  \nonumber, \\
B_\mathrm{norm} = \frac{\sinh^{-1}(e^{10}\times I)}{\sinh^{-1}(e^{10})} \times \frac{B}{I} + 0.05.
\end{eqnarray}

where $B$ are the fluxes of each pixel in the respective bands, while $B_\mathrm{norm}$ represents  
the fluxes of each pixel after scaling for each of the bands $g$, $r$ and $i$. We choose this normalization as opposed to the square-root stretch as it performs slightly better in our tests.

We then apply data augmentation to the dataset as follows:
\begin{enumerate}
\item a random shift ranging between -6 and +6 pixels in both the x and y-directions
\item a random horizontal and vertical flip, each with 50\% probability
\item a random rotation in the range [-36,36] deg
\item a random adjustment of the image contrast in the range [0.9, 1.1]
\item a random scaling of the image brightness in the range [-0.1, 0.1]
\end{enumerate}

\begin{table}
   \centering
    \caption{Summary of various test datasets}
    \label{tab:sample_summ}
       \begin{tabular}{lcccccc}
        \hline
     Dataset && Dataset  && Dataset && Dataset \\
     Name && size && Name  &&  size   \\
        \hline
    L1 (Real)       &&  42    &&  N1 (SW)   &&  2996  \\
    L2 (Real)       &&  138   &&  N2 (C21)  &&  3000  \\
    L3 (Mock-C21)   &&  3000  &&  N3 (S22)  &&  3000  \\
    L4 (Mock-J23)   &&  3000  &&  N4 (J23)  && 3000  \\
    L1+L2           &&  180   &&  N5 (SW)   &&  727   \\
    L (all)         &&  6180  &&  N (all)   &&  12723 \\
    \hline
    \end{tabular}
\end{table}

The data augmentation is applied directly to the input training and validation data at the start of the training (i.e., it does not create duplicate objects) and is not re-applied at each epoch.  The data augmentation steps can result in the transformed images containing points outside the original cutouts of the input images, so we fill these regions with zeros to maintain $64\times64$ pixel cutouts.  We tested other fill modes, including reflection, wrap, and nearest pixel and found that they gave similar results.  The Adam optimization algorithm was chosen to minimize the binary cross-entropy error function over the training data with a learning rate 0.001.  We use a batch size of 64 images.  Early stopping is used to save the best model to minimize the influence of overfitting if the network does not improve within five epochs.  We originally used a 70/15/15 train/validation/test split for both the mock lenses and the non-lenses.  However, we decided to use a set of $\sim200$ real galaxy-galaxy lenses from the SuGOHI sample combined with the 15\% of excluded non-lenses for our test sample with which we evaluate the performance of the network, so the 15\% of mock lenses was returned to the training sample.  This effectively results in a 85/15 train/validation split for the mock lenses and a 70/15/15 train/validation/test split for the non-lenses.

\begin{table*}
    \centering
    \begin{center}
    \caption{Comparison of performance for different combinations of test datasets.}
    \label{tab:perf_comp}
    \begin{tabular}{lllllllll}
        \hline
        Test datasets & \multicolumn{4}{c}{AUROC} & \multicolumn{4}{c}{F1$_{\rm thresh}$}  \\
        \cmidrule(ll){2-5}\cmidrule(ll){6-9}
         & C21 & S22 & J23 & I24 & C21 & S22 & J23 & I24  \\
        \hline
        a) L1+L2$-$N1 & 0.97 &  0.98 &  0.86 &  0.80 &  0.77 &  0.76 &  0.55 &  0.45 \\
        b) L1+L2$-$N2 & 0.97 &  0.91 &  0.90 &  0.80 &  0.74 &  0.53 &  0.61 &  0.47 \\
        c) L1+L2$-$N3 & 0.97 &  0.99 &  0.88 &  0.83 &  0.78 &  0.77 &  0.62 &  0.48 \\
        d) L1+L2$-$N4 & 0.94 &  0.96 &  0.97 &  0.87 &  0.68 &  0.72 &  0.74 &  0.51 \\
        e) L1+L2$-$N5 & 0.96 &  0.90 &  0.84 &  0.79 &  0.77 &  0.70 &  0.69 &  0.52 \\
        f) L1+L2$-$N (all) & 0.96 &  0.96 &  0.90 &  0.82 &  0.59 &  0.47 &  0.35 &  0.37 \\
        g) L3$-$N2 & 1.00 &  0.99 &  0.90 &  0.71 &  0.99 &  0.94 &  0.70 &  0.25 \\
        h) L3$-$N3 & 1.00 &  1.00 &  0.88 &  0.76 &  1.00 &  0.96 &  0.70 &  0.25 \\
        i) L3$-$N4 & 1.00 &  1.00 &  0.96 &  0.82 &  0.99 &  0.96 &  0.71 &  0.25 \\
        j) L4$-$N2 & 0.80 &  0.70 &  0.98 &  0.99 &  0.18 &  0.30 &  0.95 &  0.95 \\
        k) L4$-$N4 & 0.65 &  0.81 &  0.99 &  0.99 &  0.17 &  0.31 &  0.97 &  0.96 \\
        l) L (all)$-$N (all) & 0.87 &  0.91 &  0.94 &  0.87 &  0.70 &  0.70 &  0.82 &  0.68 \\
        \hline
    \end{tabular}
    \end{center}
\end{table*}

\section{Construction of the common test samples}
\label{sec:comm}
Here, we describe the various real and simulated lens and non-lens samples used in constructing the common test datasets for the networks to be compared systematically and to benchmark their performances. The participating teams had agreed that all of the HSC data from the GAMA09H field be reserved for testing and comparison of various networks. A summary of sample sizes of the various datasets are given in Table~\ref{tab:sample_summ} and their further details are given in the following. 

\begin{figure*}
    \centering
	\includegraphics[width=1.0\textwidth]{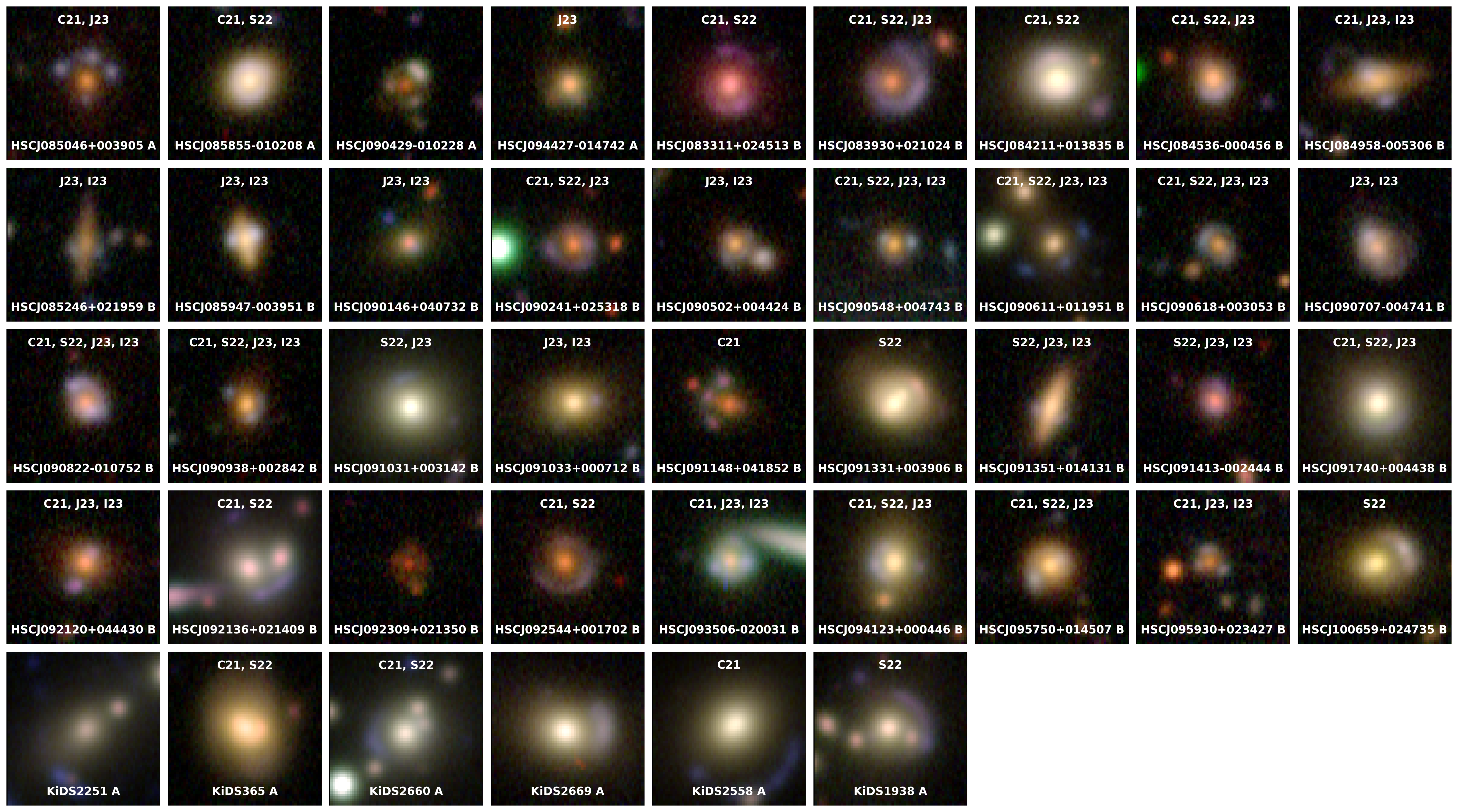}
    \caption{Mosaic of grade A and B galaxy-galaxy lenses and lens candidates in dataset L1. We list the neural 
    networks that successfully classify them as lenses, for the score threshold assumed in the corresponding 
    lens search papers.}
    \label{fig:mosaic_setL1}
\end{figure*}

\begin{figure*}
    \centering
	\includegraphics[width=1.0\textwidth]{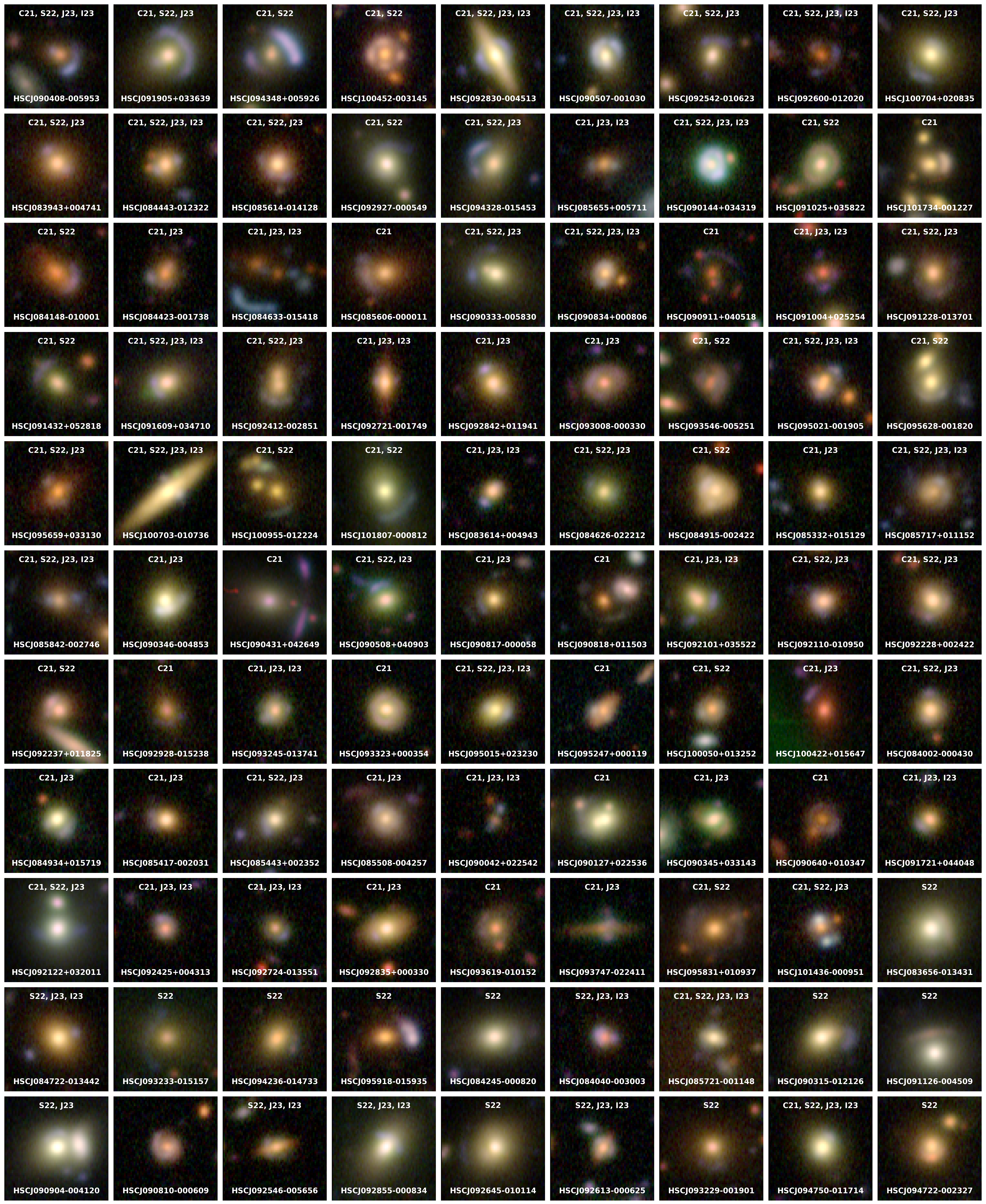}
    \caption{Mosaic of grade A and B galaxy-galaxy lenses and lens candidates in dataset L2, obtained from our
    four machine learning searches in HSC PDR2 images. We list the neural networks that successfully classify 
    these cutouts as lenses, for the score thresholds used in the corresponding papers.}
    \label{fig:mosaic_setL2a}
\end{figure*}

\begin{figure*}
    \centering
	\includegraphics[width=1.0\textwidth]{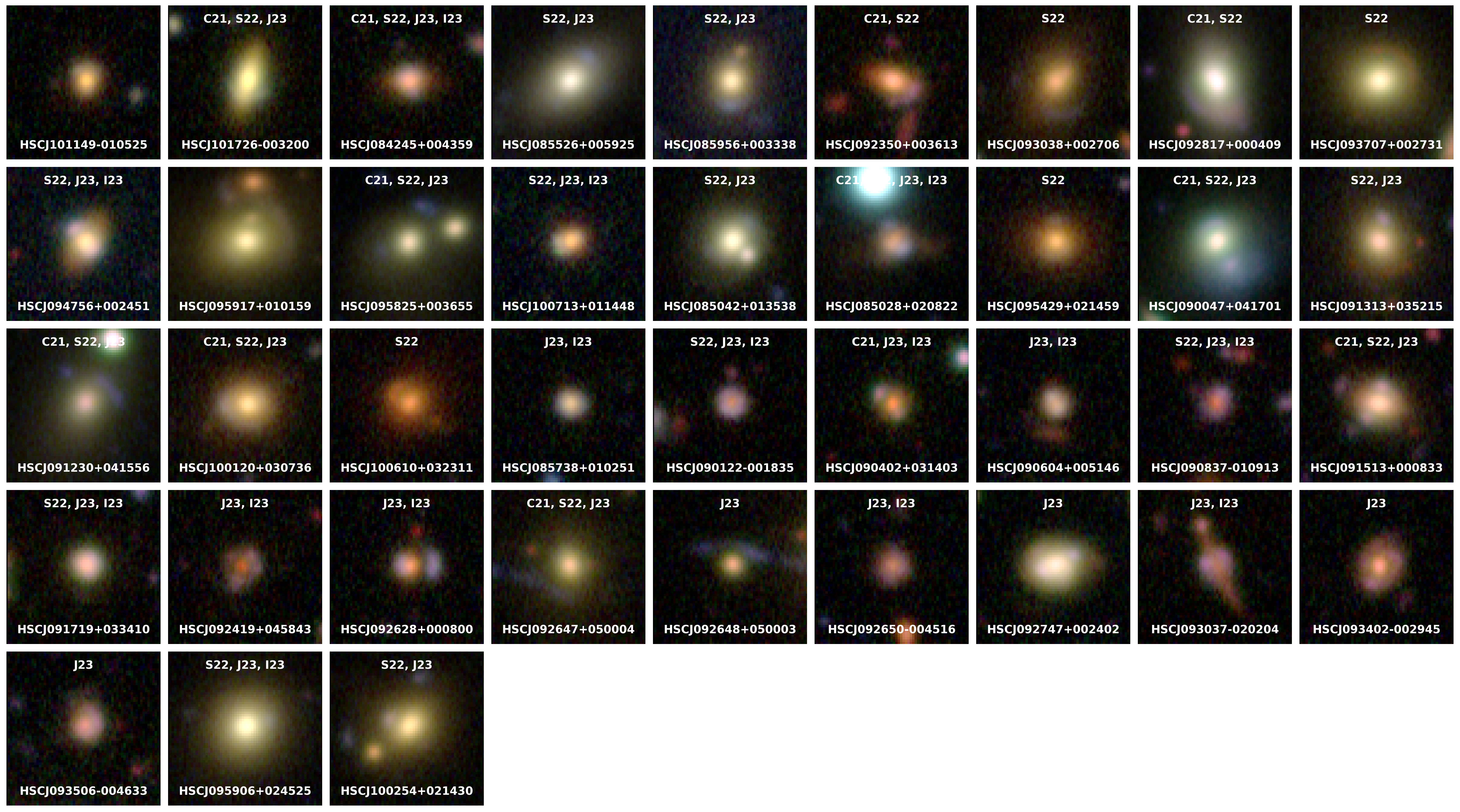}
    \caption{(Continued).}
    \label{fig:mosaic_setL2b}
\end{figure*}

\subsection{Known galaxy-scale lenses - L1}
Each network is tested on an observational dataset of 42 galaxy-scale strong lenses in GAMA09H that have been 
either spectroscopically-confirmed or listed as high-quality candidates. First, we use all systems listed as galaxy-galaxy 
systems with grade A or B in SuGOHI papers \citep{sonnenfeld18,sonnenfeld20,wong18,jaelani20}, which corresponds to four 
grade A  and 32 grade B. These lenses were found in HSC Wide imaging from a range of data releases up to PDR2 either with
Yattalens, an arc-finder combining lens light subtraction and lens modeling, or with crowdsourcing. All high-quality 
candidates were also validated by experts. Second, we consider the galaxy-scale lens candidates identified in GAMA09H with 
deep learning classification of images from Data Release 4 of the Kilo-Degree Survey \citep[LinKS,][]{petrillo17,petrillo19}. 
We only consider the subset classified as highest-quality, with a visual score larger than 28 in the grading scheme adopted 
by the authors.

In summary, strong lenses in dataset L1 have been found either via non-machine-learning techniques applied to HSC multiband
imaging, or via supervised CNNs applied to KiDS imaging, but none has been identified by neural networks from the HSC Wide
images we are testing the networks on. They cover a large variety of multiple-image configurations and angular separations, 
as well as various source over lens flux ratios (see Fig.~\ref{fig:mosaic_setL1}).

\subsection{Lens candidates from our own networks - L2}
The dataset L2 contains lens candidates found in HSC PDR2 images of GAMA09H with three of our networks\footnote{I24 network is yet to be run on the entire HSC footprint and does not yet have a corresponding sample of lens candidates.}. After removing duplicates and 
galaxy-scale systems part of dataset L1, we obtained 138 grade A or B candidates with visual grades $\geq$1.5. A small fraction
of these candidates are already published in the literature, including a few group-scale lenses from SuGOHI that were not
considered for dataset L1. A total of 80, 79, and 36 systems were originally selected by the neural networks from 
C21, S22, and J23, respectively. We noticed that reclassifying these 138 strong lens 
candidates results in the recovery of 94, 89, and 93 systems for C21, S22, and J23,
respectively (see Fig.~\ref{fig:mosaic_setL2a} and \ref{fig:mosaic_setL2b}). This discrepancy is likely mainly coming 
from different selections of the parent samples and different CNN selection functions, and partly from the uncertainties 
inherent to the human inspection process. For instance, of the 58/138 systems in L2 and out of the grade A and B sample 
from C21, 8 were discarded from the selection of the parent sample, 42 were excluded by the ResNet,
and only 8 were assigned low grades $<$1.5 by visual inspection. 

\subsection{Mock lenses from Canameras et al. - L3}
The dataset L3 includes 3000 mock lenses generated following the methodology of C21 except the GAMA09H field which was excluded 
during training. In short, the simulations follow the procedure described in \citet{schuldt21} and \citet{canameras20} by 
co-adding lensed sources to HSC Wide images of LRGs in GAMA09H with SDSS DR16 spectroscopy. We used the spectrosocpic redshifts $z_{\rm spec}$ 
and velocity dispersions $v_{\rm disp}$ from SDSS as a proxy to model the lens mass distributions with Singular Isothermal Ellipsoids (SIE), 
and we deduced the SIE centroids, position angles, and orientations from the $i$-band light profiles with some scatter.
Multiband cutouts of high-redshift background sources were taken from the {\it Hubble} Ultra-Deep Field \citep[HUDF,][]{inami17},
with neighbouring galaxies masked with SExtractor \citep{bertin96}, and random flux boosts applied to all three bands
in order to ensure all arcs are detectable in the image plane. Given the lens and source properties, pairs were matched
iteratively to ensure a flat Einstein radius distribution between 0.75\arcsec\ and 2.5\arcsec. During this process, we 
gave more weight to lens galaxy at $z>0.7$ in order to boost the fraction of distant lenses relative to the input lens (LRG) redshift 
distribution peaking at $z \sim 0.4$--0.6. The lens galaxies were used up to four times, in which they were rotated by k $\pi$/2, 
to ensure they appeared only once with a given orientation.

For each lens system, the HUDF-selected source was randomly centered at positions with $\mu \geq 5$ in the source plane, lensed to the image plane with {\tt GLEE} \citep{suyu10,suyu12}. The lensed images were convolved with the subsampled HSC PSF model, scaled to the HSC
pixel size, photometric zero point accounted for, degraded with Poisson noise, and finally coadded with the lens galaxy image cutout. New source 
positions were drawn until the lensed arcs reached ${\rm S/N \geq 5}$ with respect to the local sky background around the 
lens galaxy, and until their peak flux exceeded the lens brightness at the peak image position, either in $g$ or $i$ band. 
These thresholds discard all mocks with faint multiple images or strong lens-source blending. The process also guarantees 
that resulting mocks reproduce the local variations in  depth and seeing, and include line-of-sight objects and artefacts.

\subsection{Mock lenses from Jaelani et al. - L4}
The dataset L4 is a sample of 3000 mock lenses from GAMA09H generated by J23 and excluded during their training.  These mocks are hybrid in nature, that is, simulated lensed features are superposed on real images of HSC galaxies. We use the {\sc simct}\footnote{\url{https://github.com/anumore/SIMCT}} pipeline \citep[see][for details]{More16} for this purpose. Starting with a catalog of massive galaxies, we make use of the photometric redshifts and magnitudes to determine the luminosity of the galaxies. Using standard relations, we obtain the velocity dispersion and assuming mass follows light, we characterise the lens mass model for each potential lens galaxy. The background galaxies are drawn from luminosity functions, colors are taken from real galaxy catalogs and light profile is parameterised with a Sersic model. Lensed arc-like images are generated of the Sersic model for the background source which are then merged with the $griz$ images of the respective lens galaxy facilitiating realistic noise, image quality, lens environments and so on. Finally, only those lens galaxies which have sufficient lensing optical depth and lensed images that meet certain detectability criteria are retained in the mock sample. 

\subsection{Random non-lenses - N1}
We randomly selected 2996 non-lenses from the GAMA09H field which were classified by the Space Warps citizens as low lensing 
probability candidates. We ensure that none of the SuGOHI lenses or the recently-published grade A and B strong lens 
candidates are part of this sample. 

\subsection{Selection of non-lenses following Canameras et al. - N2}
This dataset includes 3000 real galaxies from GAMA09H selected with the same recipe as non-lenses for training the network 
in C21. Galaxies are selected from the HSC Wide DR2 catalog to match four different classes. 
First, a total of 33\% spiral galaxies from \citet{tadaki20} with $i$-band Kron radii below 2\arcsec. This size cut 
is intended to exclude the brightest, most extended galaxies in the input catalog to focus on spirals with arms 
located at similar angular separation as multiple images of galaxy-scale lenses. Second, 27\% LRGs from the input 
sample of the lens simulation pipeline, namely isolated LRGs from dataset L2 without lensed arcs. Third, 6\% compact 
galaxy groups selected from \citet{wen12}, with at least four galaxies, falling within the HSC cutout. Fourth, 
33\% random galaxies with r$_{\rm Kron} < 23$~mag.

\subsection{Selection of non-lenses following Shu et al. - N3}
The dataset N3 includes 3000 real galaxies within the GAMA09H field that satisfy a dataset of criteria defined in Section 2 of 
S22, which were used to construct the non-lens examples for training, validation, and testing purposes. When 
selecting, we made sure that none of the 3000 galaxies was included for training or validation in S22 and none 
of them were reported as a strong-lens system or candidate according to a strong lens compilation built by S22.

\subsection{Selection of non-lenses following Jaelani et al. - N4}
The dataset N4 includes 3000 real galaxies from GAMA09H following the similar selection as non-lenses in J23. We selected non-lens 
objects for the negatives, which containing: 40\% galaxies that are randomly selected with photometric redshift range between 0.2 
and 1.2, and $i-$magnitude $<28$; 30\% (tricky or merge) spiral galaxies from \cite{tadaki20} combined with visual investigation; 
25\% galaxy groups or "crowded" galaxies like LRG + egde-on galaxy (or arc like feature); and 5\% dual point-like.

\subsection{Tricky non-lenses - N5}
The dataset N5 comprise of a sample of 727 non-lenses in GAMA09H from SpaceWarps (found by YattaLens and visually classified as 
FP; used for training Citizens), after excluding any overlap with SuGOHI or with recently-published grade A or B strong lens 
candidates.

\section{Metrics used in comparison analysis}
\label{sec:metrics}
\begin{figure*}
    \centering
	\includegraphics[width=1.\textwidth]{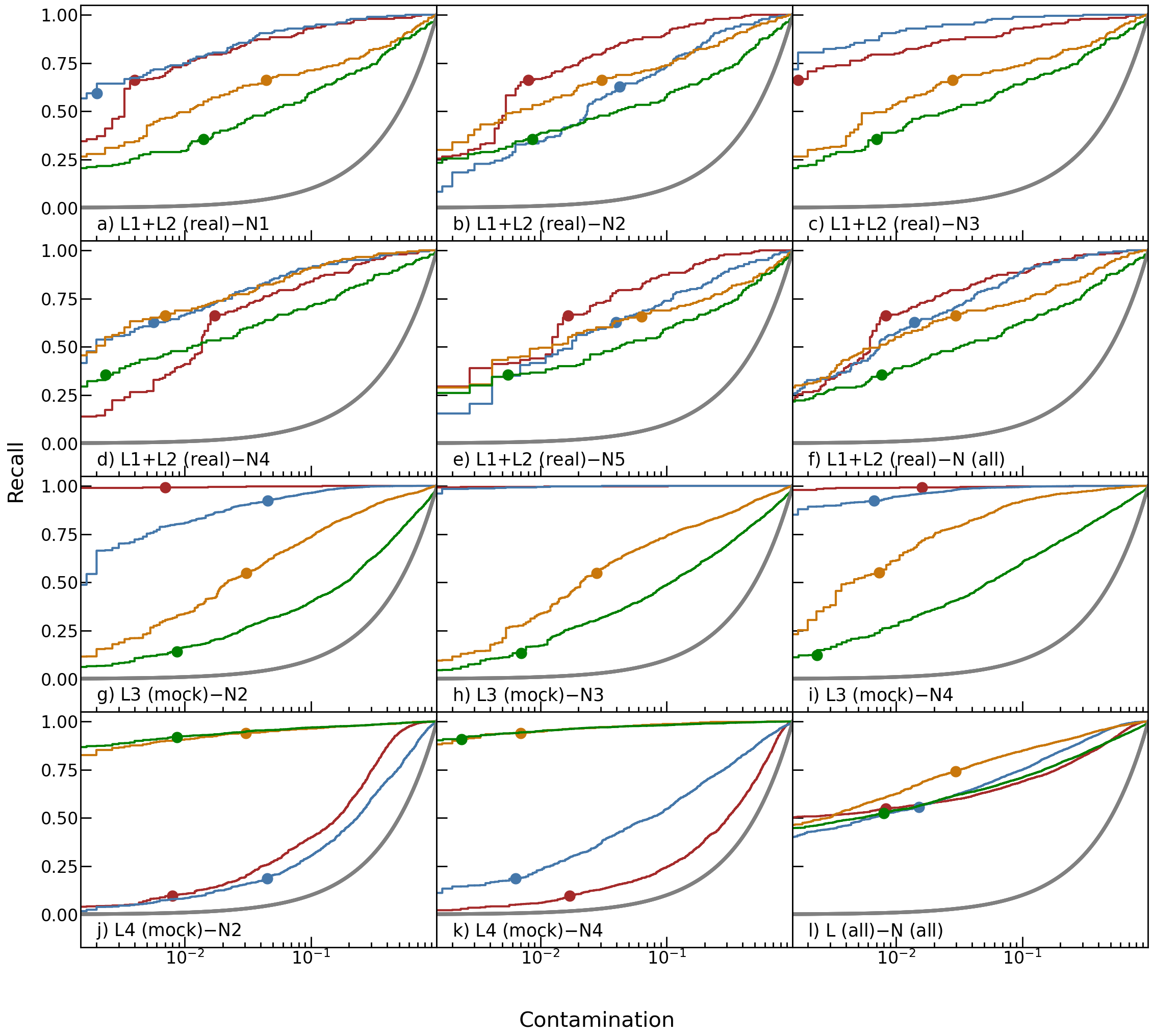}
    \caption{ROC curves for the networks presented in C21 (maroon), S22 (blue), J23 (orange), and I24 (green). The first and second rows focus on test datasets drawn from observations. The five first panels from 
    top-left measure recall with real SuGOHI and KiDS lenses and lens candidates from dataset L1+L2, and derive contamination from
    various combinations of non-lenses in datasets N1, N2, N3, N4, and N5. The last panel in the second row combines all lenses 
    and non-lenses with real HSC images (excluding the mock lenses in datasets L3 and L4), after removing duplicates. Panels in 
    the bottom two rows focus on positive and negative examples used for training networks in C21, S22, and J23. Sections with 
    $<$5 objects are masked to account for the variations in sample sizes. The threshold scores defined in each paper are 
    indicated by dots. The thick grey lines show a random classifier.}
    \label{fig:roc_curves}
\end{figure*}

The performance of each network is evaluated with a range of metrics, and various 
combinations of test datasets. Firstly, the Receiver Operating Characteristic (ROC) curves 
are computed by varying the network thresholds between 0 and 1, as shown in Fig.~\ref{fig:roc_curves},
using the following definitions of the true positive rate (TPR or recall) and false positive 
rate (FPR or contamination):
\begin{equation}
\rm TPR = \frac{TP}{TP+FN}; FPR = \frac{FP}{FP+TN}\,.
\end{equation}

This allows us to deduce the classical metrics for binary classification problems that are used in the previous lens finding challenges \citep{metcalf19} and other studies \citep{schaefer18,cheng20}, namely, the Area Under the ROC (AUROC).
Computing these quantities for the current HSC test datasets, including a wide range of spirals, rings, 
mergers, and other types of non-lens galaxies, allows us to compare the network classification performances with previous challenges focusing on less representative test samples drawn from simulations. 

\begin{figure*}
 \centering
   \includegraphics[scale=0.22]{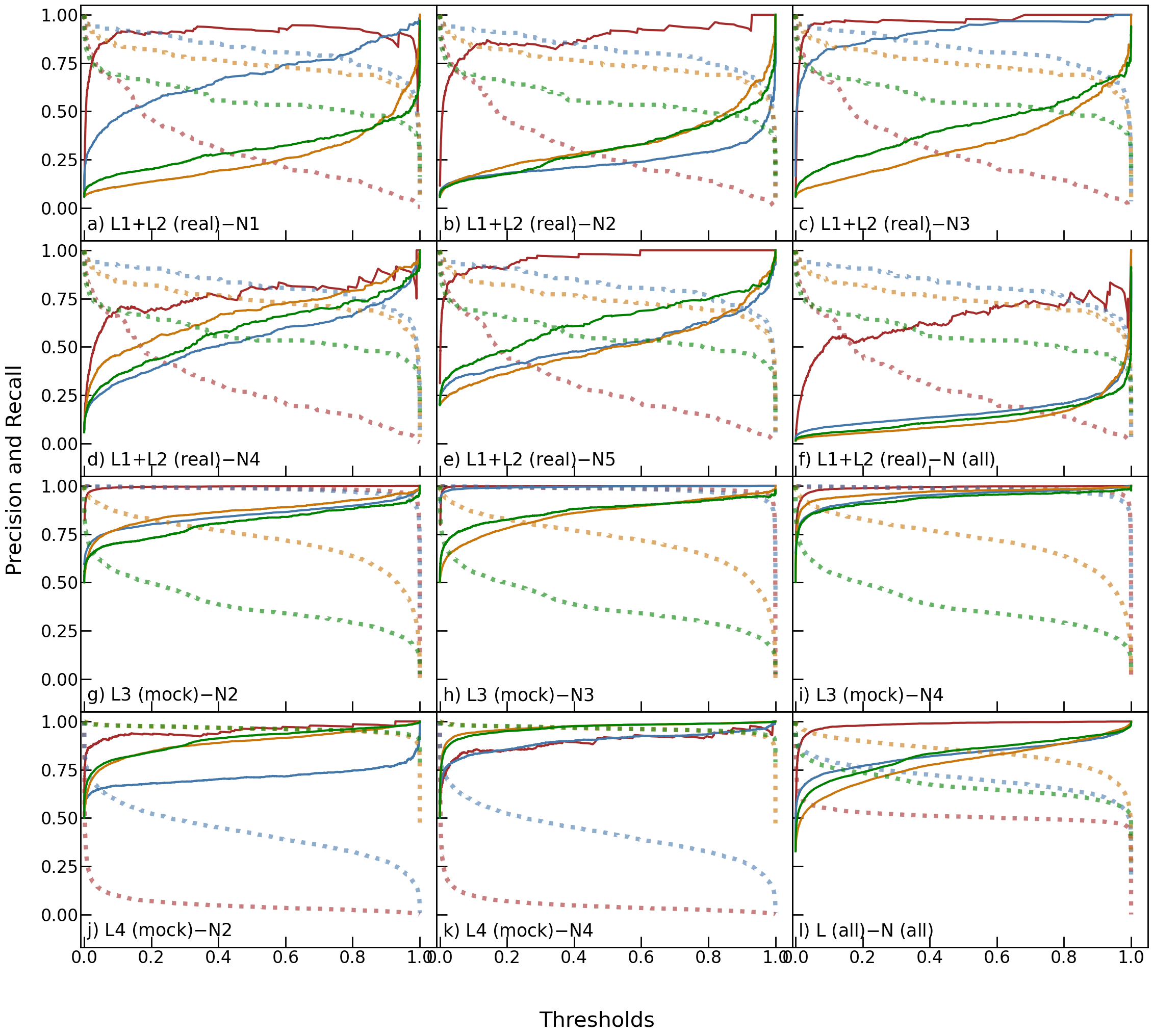}
   \includegraphics[scale=0.22]{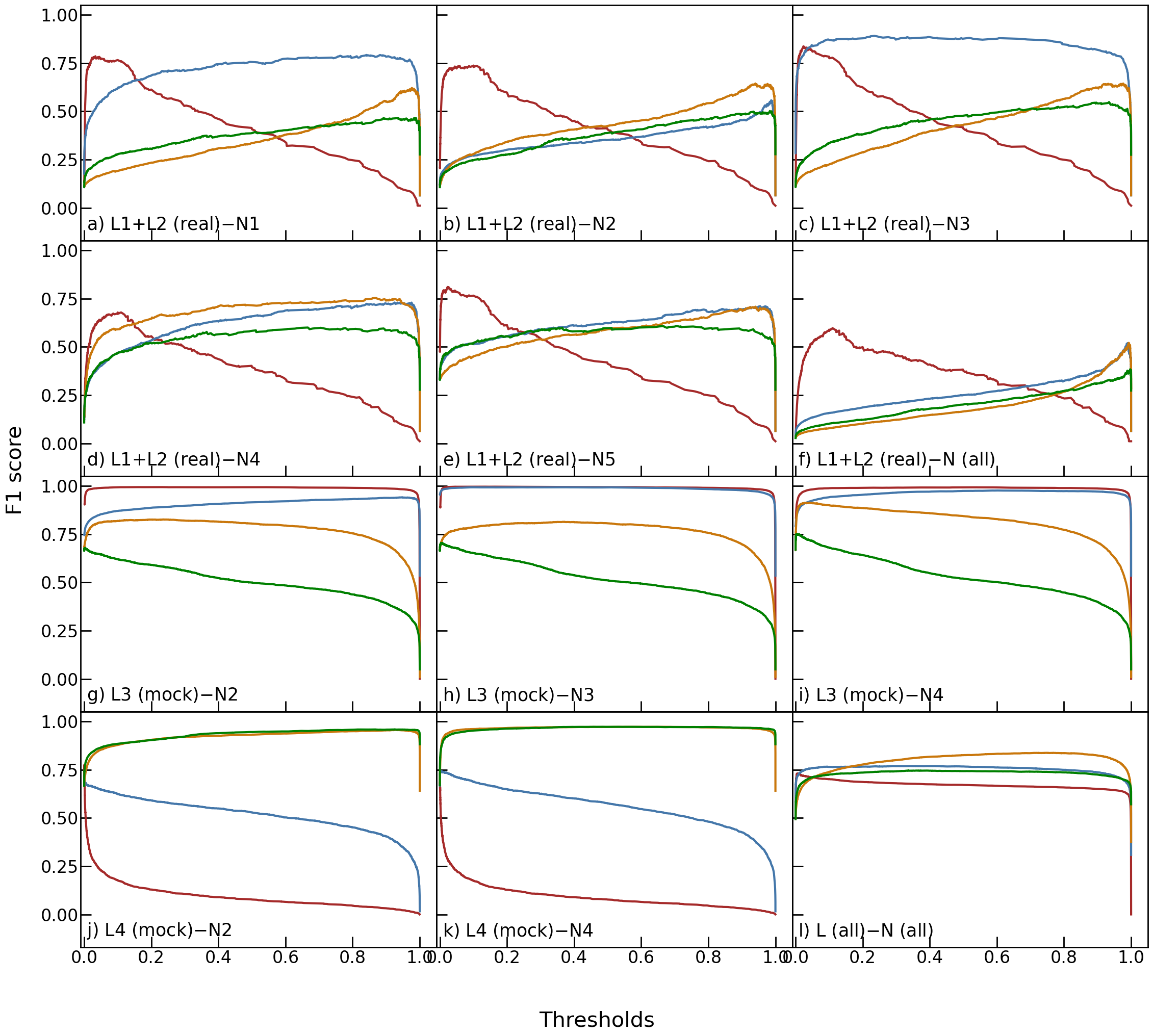}
     \caption{Performance metrics as a function of network threshold for the same combination of datasets as shown in Fig.~\ref{fig:roc_curves}. {\it Top:} Precision (solid) and Recall (dashed) curves and {\it Bottom:} F1 scores for the four networks by C21 (maroon), S22 (blue), J23 (orange), and I24 (green).}
    \label{fig:metrics}
\end{figure*}

In Fig.~\ref{fig:metrics}, we show an additional metric called the F1 score as a function of the threshold score. We use the standard definitions of F1 score, precision and recall (i.e. TPR) as follows:
\begin{equation}
\rm   F1 = 2 * \frac{precision * recall}{precision+recall} ; \,\, \,\, precision =\frac{TP}{TP+FP}\,.
\end{equation}

While an ideal network would have both high precision and high recall, the networks, in practice, tend to perform better on only one of them while compromising the other. The F1 score allows one to assess the accuracy of the network by combining precision and recall. It is defined to result in a high value when both the precision and the recall are high. 

We also compute the performance for different combinations of test datasets. We start exclusively with test datasets drawn from 
observations by measuring recall from SuGOHI and KiDS lenses and lens candidates from dataset L1, and deriving contamination from 
non-lenses in datasets N1 to N5. We then combine all non-lenses together, we include lens candidates from dataset L2, and we 
estimate the performance jointly for all real and simulated lenses, and for all non-lens galaxies. Finally, we focus on datasets 
L3/N2, L3/N3, and L4/N4 that mimick the positive and negative examples used for training networks in C21, S22, and J23, 
respectively. These various combinations of test datasets range from 40$-$6200 positive examples and 700$-$12700 negatives (see 
Table~\ref{tab:sample_summ}).

\section{Results of Comparison}
\label{sec:results}
We have characterised various networks using the different metrics listed in the previous section. We have two networks based on the ResNet architecture i.e. C21 and S22. Their respective training sets are L3$+$N2 and L3$+$N3. We have two other networks based on conventional CNN architecture i.e. J23 and I24 which are using training sets L4$+$N4 with minor differences in the augmentation. 

\subsection{Comparison of different networks}
\label{subsec:comp_networks}
The Fig.~\ref{fig:roc_curves} shows the ROC curves for different combinations of test datasets in each panel where the random classifier (dark grey) always serves as a reference. The real (candidate) lens sample (L1$+$L2) combined with each non-lens sample, N1 to N5, are shown in panels "a" to "e", respectively. The L1$+$L2 combined with N$-$all is shown in panel "f". By and large, the ResNet-based and L3-trained networks C21 (maroon) and S22 (blue) have a better performance. From panels "g" to "k", we interchange both the simulated lenses and the non-lenses from each of the training sets to better understand the sensitivity and robustness of each of the four networks. Needless to say each network performs the best when trained on its own training dataset. The C21 and S22 also perform well when tested on the non-lens sample, N4, of J23 (see panel "i"). However, their ROC curves worsen when tested specifically on the L4 (simulated lens) dataset (panels "j" and "k") regardless of the type of the non-lens sample (N3 is skipped for simplicity). 

A similar behaviour is noted for J23 (orange) and I24 (green) networks. In the final panel "l" where all (simulated and real) lenses are combined with all non-lenses,  the J23 (orange) network performs better than the rest, even if marginally so. The main reason being the scatter in the ROCs of J23 across different dataset combinations is smaller than the other networks providing a relatively stable performance. While the above conclusions are also evident from a quantitative result based on the AUROCs reported in Table~\ref{tab:perf_comp}, we want to emphasize that the differences between the AUROCs for various networks across different dataset combinations are fairly small (particularly for rows f) and l)). In fact, the Fig.~\ref{fig:roc_curves} is made linear$-$log in order to be able to clearly see the differences between the different networks.

Next, we show the Precision (solid curves -top half), Recall (dotted curves -top half) and F1 score (bottom half) in the Fig.~\ref{fig:metrics} as a function of thresholds for the same combination of datasets as in the previous figure with the ROCs. Since F1 combines both precision and recall of a network shown for varying thresholds, it allows us to compare the overall accuracy of various networks in a more comprehensive manner across different datasets.     
In the panel "f" of top half of Fig.~\ref{fig:metrics} when analysing real lenses and non-lenses, we see that the precision of S22, J23 and I24 are all comparable and relatively poor as compared to C21 which has much higher precision. In terms of recall, S22 shows the best performance with J23 following closely and I24 and C21 come next in that order. In panel "l" which now also includes simulated lenses, the performance improves quantitatively but the same qualitative trends are seen as panel "f".  

In the bottom half of Fig.~\ref{fig:metrics}, we find that apart from C21, all of the other networks show similar rising or nearly constant trend in the F1 score with increasing thresholds. Thus, a low threshold for C21 and high thresholds for other networks are expected to give better network accuracy and performance. It is not a surprise then that the thresholds chosen for the respective network by C21, S22, J23 and I24 are about 0.1, 0.97, 0.99 and 0.9, respectively. In spite of being trained on the same simulated lenses, the distinction between C21 and S22 networks becomes more apparent in panels "a" to "f" when tested on real lenses combined with different kinds of non-lenses. The F1 score curve of S22 shows moderate to marginally better performance in most of these panels followed by J23 and then I24. 

As before, the networks trained on their own simulated lenses (L3 for C21 and S22 whereas L4 for J23 and I24, see panels "g" to "k") produce superior F1 score at all thresholds. On the all lens - all non-lens samples (panel "l"), the J23 network performs better than others. These results highlight a general and prevalent issue of over-fitting of networks to specific training samples which then result in an unknown and usually, poorly characterised performance on real data. 

We also report the F1 score for the aforementioned specific thresholds in Table~\ref{tab:perf_comp}. These thresholds, chosen by each team, are applied when conducting the actual lens search on the entire HSC survey data (except for I24 which has not been applied to the entire HSC data yet). We note similar trends as before. The F1$_{\rm thresh}$ score of C21 is higher on most dataset combinations except when trained on J23 dataset. However, when all lenses and all non-lenses are combined all networks have comparable F1 scores with J23 leading marginally owing to a smaller scatter overall across different datasets.

\subsection{Comparison of networks when trained on interchanged datasets} \label{subsec:train_interchanged}
To further understand the possible causes of the different trends seen in previous section, we decide to perform a number of tests among the various networks in which we keep each individual network architecture the same, but train on the training data initially used by the other teams. These tests will help us to assess the the role of training sample on the performance of the networks.

\begin{figure*}
    \centering
	\includegraphics[width=1.0\textwidth]{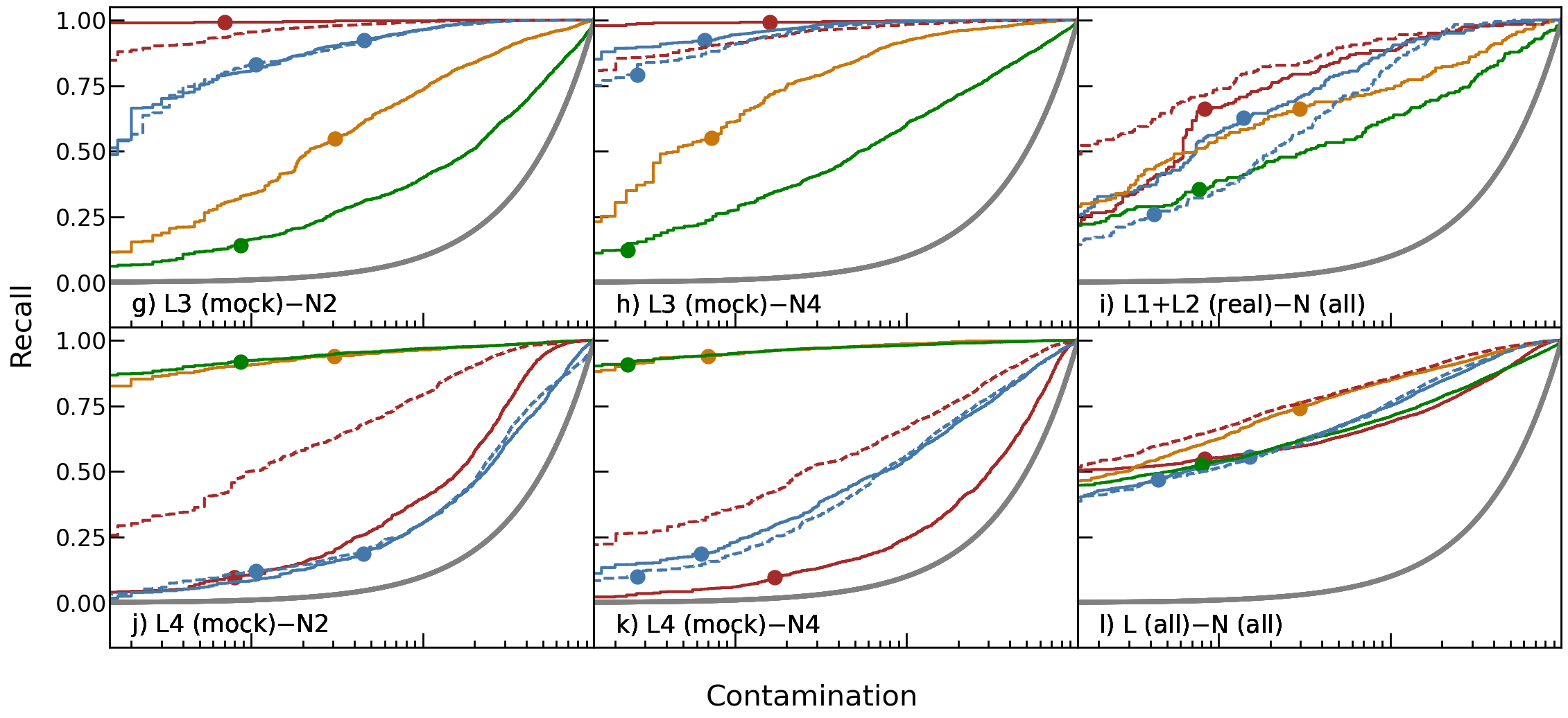}
    \caption{Comparison of performance for the network from I24 when trained on different training data samples. For simplicity, only a subset of cases are shown. Dashed curves show the I24 CNN trained on the dataset from C21 (maroon) and from S22 (blue). For reference, solid lines reproduce the original curves from Fig.~\ref{fig:roc_curves}, namely for the C21 (maroon), S22 (blue) and J23 (orange) networks trained on the corresponding datasets. Note that I24 (green solid curve) was originally trained on J23 dataset.}
    \label{fig:roc_yuichi_swap}
\end{figure*}

The I24 network, which initially was trained on the J23 training sample, is subsequently trained on the C21 and S22 training samples with no modifications to the network architecture. The preprocessing steps, including the SDSS normalization and data augmentation described in Section~\ref{sec:net}, are kept the same and applied to the new training samples. The results of these tests are shown in Table~\ref{tab:training_sample_test} and Figure~\ref{fig:roc_yuichi_swap} where the dashed curves represent the new I24 performance and the solid curves of the four original networks are also shown for reference. Here, curves of same color will have same training datasets.  

\begin{figure*}
 \centering
   \includegraphics[width=1.\textwidth]{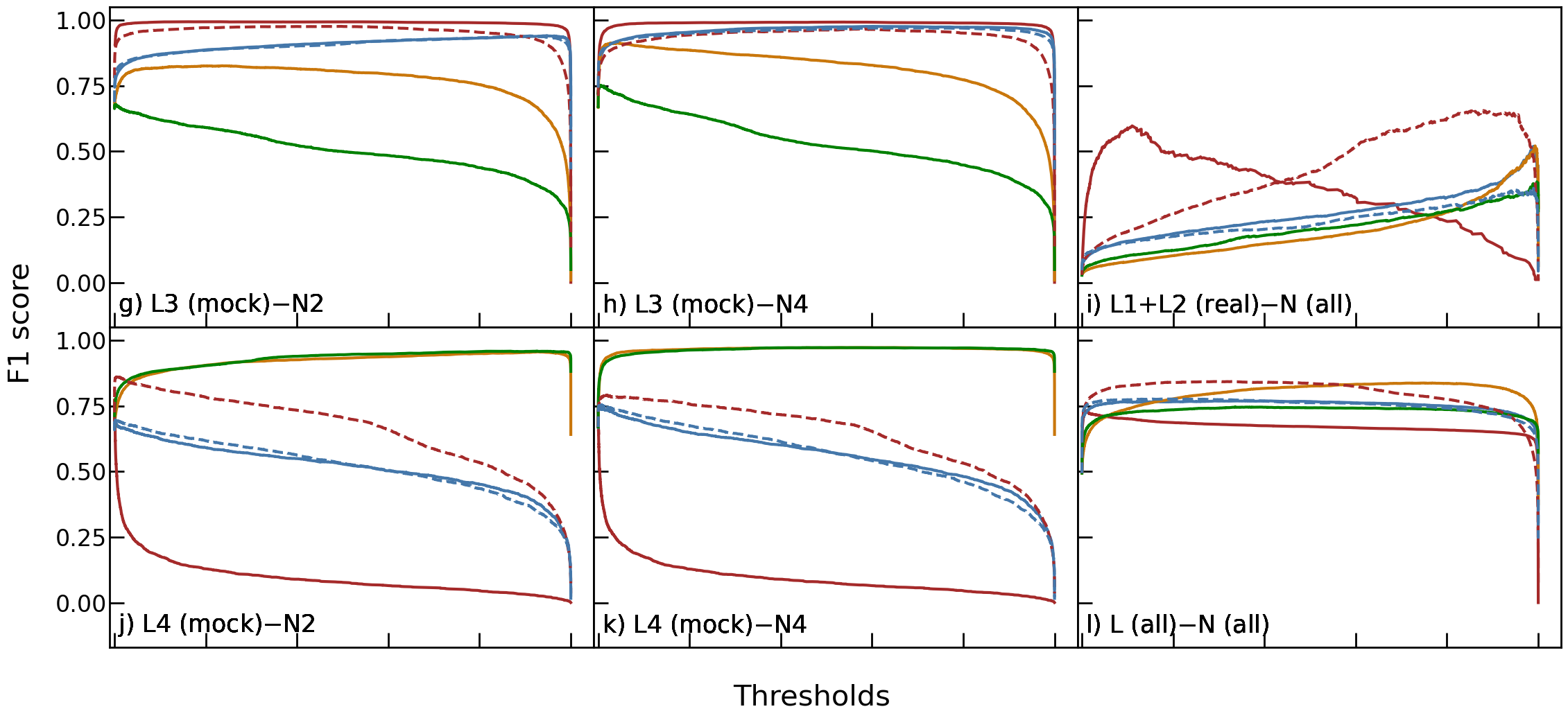}
     \caption{Performance metrics F1 score of I24 as a function of threshold for the same combination of datasets as shown in Fig.~\ref{fig:roc_yuichi_swap}. As before, the dashed curves show the modified F1 score performance of I24 when trained on C21 dataset (maroon) and S22 dataset (blue).}
    \label{fig:perf_yuichi_swap}
\end{figure*}

Based on the metrics, the I24 network performs better when trained on the C21 (maroon dashed) and S22 (blue dashed) datasets than when trained on the J23 datasets (green solid), even though
the network architecture is unchanged. In fact, comparing the dashed and solid maroon curves in Fig.~\ref{fig:roc_yuichi_swap} indicates that, for most combinations of test datasets, the I24 network even performs comparably or even better than the C21 network (see panels "i","l"). Furthermore, the F1 score as a function of the threshold (Figure~\ref{fig:perf_yuichi_swap}) too shows that I24 improves with training on the C21 and S22 datasets (dashed curves with respect to solid green, see panels "g","h","i","l") and it outperforms C21 (dashed with respect to solid, panels "i","j", "k" and "l").  

\begin{figure*}
    \centering
   \includegraphics[width=1.0\textwidth]{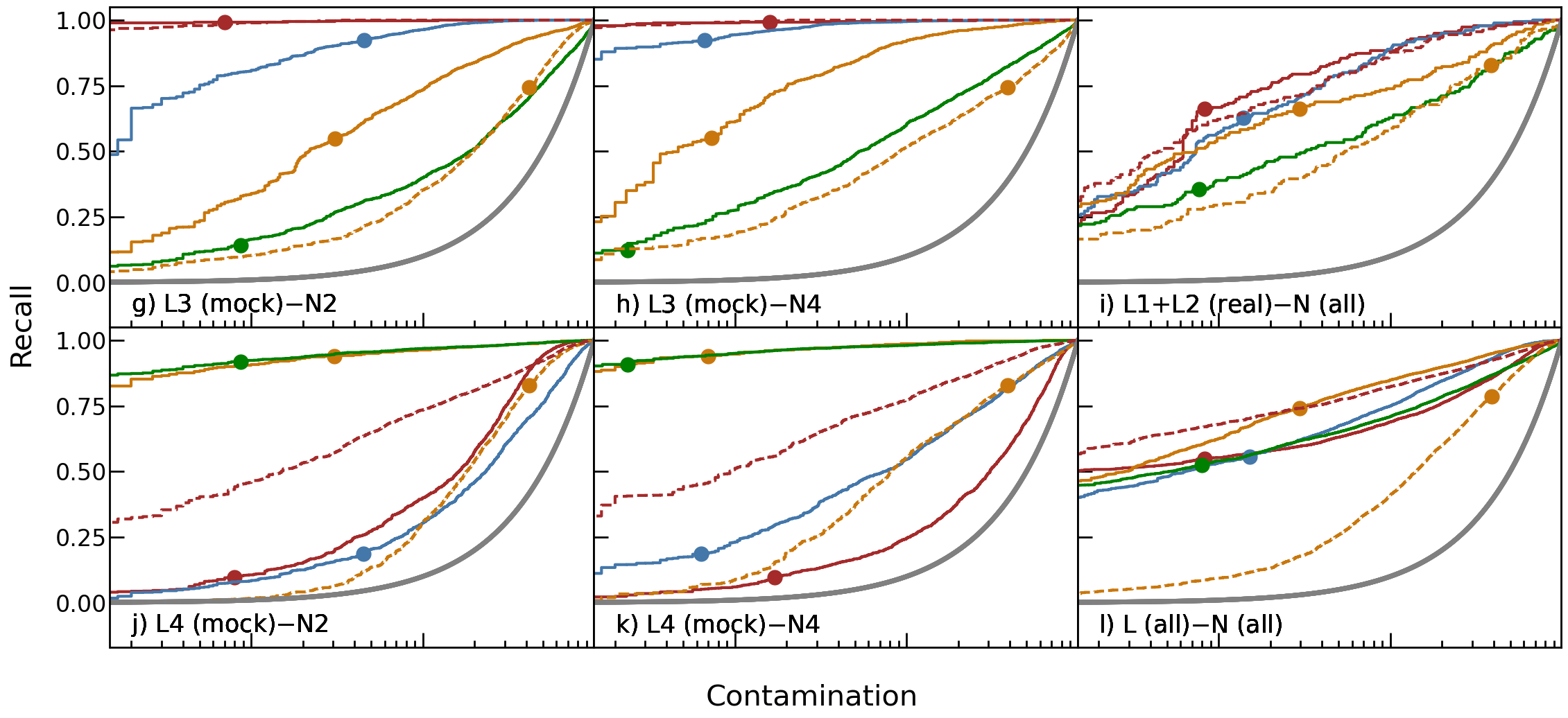}
    \caption{Comparison of performance for the network from C21 when trained on J23 (orange dashed) and the one from J23 when trained on C21 (maroon dashed). For reference, solid curves reproduce the original ROC from Fig.~\ref{fig:roc_curves}, namely for the C21 (maroon), S22 (blue), J23 (orange), and I24 (green) networks.}
    \label{fig:roc_raoul_anton_swap}
\end{figure*}

Encouraged by these results, we decide to perform a similar exercise with J23 and C21 networks by interchanging their training datasets. When the CNN of J23  is trained on the C21 dataset, the Fig.~\ref{fig:roc_raoul_anton_swap} shows a similar improvement in the performance of J23 (maroon dashed curve) compared to the original CNN trained on its own dataset (orange solid curve).
Interestingly, the performance of the C21 network when trained on the J23 dataset becomes substantially worse as is evident from the orange-dashed curve in Fig.~\ref{fig:roc_raoul_anton_swap} with respect to the original Resnet of C21 trained on its own dataset (maroon solid curve, panels "i" and "l"). 

These tests reinforce the fact that the training sample has a significant impact on the performance of the network. Also, the CNNs (I24 and J23) perform better on the C21 training dataset wherein the parent galaxy catalog comes from broader selection cuts (see Sec~2.1 of C21). It will be worth investigating in the future if the differences in the selection of the parent catalogs itself is the cause of these improvements which is beyond the scope of the current work.

\begin{table}
    \centering
    \caption{Comparison of performance for the network from I24 when trained on different training data samples. The original network
    I24 CNN was trained on the J23 dataset (see I24 in Table~\ref{tab:perf_comp}) which is shown here as J23 for reference. Columns labeled C21 and S22 correspond to the I24 CNN trained on their datasets, respectively.}
    \label{tab:training_sample_test}
    \begin{tabular}{l|cccccc}
        \hline
        Network & \multicolumn{6}{c}{Training Data} \\
        & \multicolumn{3}{c}{AUROC} & \multicolumn{3}{c}{F1$_{\rm thresh}$}\\
        \cmidrule(lr){2-4}\cmidrule(lr){5-7}
           & C21 & S22 & J23 & C21 & S22 & J23  \\
        \hline
        g) L3(mock)$-$N2 & 1.00 &  0.99 &  0.71 &  0.74 &  0.90 &  0.25\\
        h) L3(mock)$-$N4 & 0.99 &  1.00 &  0.82 &  0.74 &  0.91 &  0.25\\
        i) L1+L2(real)$-$N(all) & 0.97 &  0.95 &  0.82 &  0.36 &  0.33 &  0.37\\
        j) L4(mock)$-$N2 & 0.94 &  0.70 &  0.99 &  0.21 &  0.21 &  0.95\\
        k) L4(mock)$-$N4 & 0.87 &  0.82 &  0.99 &  0.21 &  0.21 &  0.96\\
        l) L(all)$-$N(all) & 0.95 &  0.92 &  0.87 &  0.52 &  0.63 &  0.68\\
                \hline
    \end{tabular}
\end{table}

\subsection{Qualitative comparison of performance on SuGOHI lenses}
We also make a qualitative comparison between the output of different networks. To illustrate this, we plot in Fig.~\ref{fig:mosaic_setL1} all grade A and B galaxy-galaxy lenses and 
lens candidates included in dataset L1 and we list the networks that would recover these systems based on their respective detection thresholds. Interestingly, HSCJ091331$+$003906 and HSCJ092309$+$021350, which have significantly redder arcs, are missed by most networks, as well as the grade A SuGOHI lens HSCJ090429$-$010228 that has a particularly compact and distant lens galaxy at a spectroscopic redshift of $z_{\rm lens}=0.957$ \citep{jaelani20}. These unusual lenses are likely not represented adequately in the training data, making them difficult to classify for any network.  KiDS2251 is classified as non-lens by all but one network, likely due to the particularly wide image separation placing the arc counter image outside of the cutout.  Objects such as this may be recovered if the training data were expanded to include larger cutouts, but this would increase the computation time required for training.  Other commonly-missed objects such as HSCJ100659$+$024735 and KiDS2669 appear to have a single thick arc and very faint or no counter image.  This particular failure mode is harder to understand, but may be improved by techniques that highlight the contrast of the faint counterimage compared to the lens galaxy light (e.g., I24).

\section{Summary and Conclusions}
\label{sec:conc}
We present one of the first systematic studies of comparison and benchmarking of multiple neural networks, for searching strong gravitational lenses, tested on the common datasets generated from the HSC Survey imaging. Four teams devised their own training samples selected from the HSC Survey data with some teams having partially common datasets during training. Every team refrained from validating and testing on the varied fixed test datasets comprising of known and/or simulated lenses and real non-lenses. Subsequently, the teams exchanged their training datasets to retrain their original networks and tested again on the same common test datasets to evaluate the (lack of) sensitivity of the networks to the nature of the training samples. We note that the analyses always includes non-lenses selected from real galaxy catalogs. We use standard metrics such as the ROC curves and F1 score that combines precision and recall for comparing the performances across the four networks.

Our main conclusions are - i) each network performs extremely well and better than the rest when trained and tested on their own datasets which are drawn from the same population of their own simulated lenses and real non-lenses ii) all networks seem to show comparable performance on the sample of real lenses (also combined with all simulated lenses) and non-lenses 
iii) While the C21 network has somewhat better AUROC on the combined test datasets, the J23 network is found to be more robust across the different combinations of test datasets. 
iv) when training datasets are exchanged, the CNNs (I24 and J23) give better performance on most test datasets and at times, outperform all of the original networks whereas the newly trained Resnet network (e.g. C21) tends to underperform on the various test datasets implying that the nature of training sample plays a crucial role v) Prima facie, the combination of CNNs and training dataset of C21 is found to give the most optimal performance which needs further investigation

\section*{Acknowledgements}

This research is supported in part by the Max Planck Society, and by the Excellence Cluster ORIGINS which is funded by the
Deutsche Forschungsgemeinschaft (DFG, German Research Foundation) under Germany’s Excellence Strategy – EXC- 2094 – 390783311.
This project has received funding from the European Research Council (ERC) under the European Unions Horizon 2020 research and innovation programme (LENSNOVA: grant agreement No 771776). A.T.J. is supported by the Riset Unggulan ITB 2024.

This paper is based on data collected at the Subaru Telescope and retrieved from the HSC data archive system, which 
is operated by Subaru Telescope and Astronomy Data Center at National Astronomical Observatory of Japan. The Hyper 
Suprime-Cam (HSC) collaboration includes the astronomical communities of Japan and Taiwan, and Princeton University. 
The HSC instrumentation and software were developed by the National Astronomical Observatory of Japan (NAOJ), the 
Kavli Institute for the Physics and Mathematics of the Universe (Kavli IPMU), the University of Tokyo, the High Energy 
Accelerator Research Organization (KEK), the Academia Sinica Institute for Astronomy and Astrophysics in Taiwan (ASIAA), 
and Princeton University. Funding was contributed by the FIRST program from Japanese Cabinet Office, the Ministry of 
Education, Culture, Sports, Science and Technology (MEXT), the Japan Society for the Promotion of Science (JSPS), Japan 
Science and Technology Agency (JST), the Toray Science Foundation, NAOJ, Kavli IPMU, KEK, ASIAA, and Princeton University. 
This paper makes use of software developed for the LSST. We thank the LSST Project for making their code available as 
free software at  http://dm.lsst.org.
This work is supported by JSPS KAKENHI Grant Numbers JP20K14511, JP24K07089.

\section*{Data Availability}
The various lens and non-lens test datasets are available upon reasonable request to the authors.

\bibliographystyle{mnras}
\bibliography{common_tests} 

\bsp	
\label{lastpage}
\end{document}